\documentclass[aps,prb,superscriptaddress,twocolumn,showpacs,amsmath,floatfix,citeautoscript]{revtex4}

\usepackage{graphicx}
\usepackage{float}
\usepackage{dcolumn}
\usepackage{color}
\usepackage{latexsym,bm}
\usepackage[normalem]{ulem}
\usepackage{multirow}
\usepackage{appendix}

\begin{document}

\hyphenpenalty=5000

\tolerance=1000

\title{Gradient optimization of finite projected entangled pair states}

\author{Wen-Yuan Liu}
\affiliation{CAS Key Laboratory of Quantum Information, University of Science and
  Technology of China, Hefei, Anhui, 230026, People's Republic of China}
\affiliation{Synergetic Innovation Center of Quantum Information and Quantum
  Physics, University of Science and Technology of China, Hefei, 230026, China}
  \author{Shao-Jun Dong}
\affiliation{CAS Key Laboratory of Quantum Information, University of Science and
  Technology of China, Hefei, Anhui, 230026, People's Republic of China}
\affiliation{Synergetic Innovation Center of Quantum Information and Quantum
  Physics, University of Science and Technology of China, Hefei, 230026, China}
\author{Yong-Jian Han}
\email{smhan@ustc.edu.cn}
\affiliation{CAS Key Laboratory of Quantum Information, University of Science and
  Technology of China, Hefei, Anhui, 230026,  People's Republic of China}
\affiliation{Synergetic Innovation Center of Quantum Information and Quantum
  Physics, University of Science and Technology of China, Hefei, 230026, China}
\author{Guang-Can Guo}
\affiliation{CAS Key Laboratory of Quantum Information, University of Science and
  Technology of China, Hefei, Anhui, 230026,  People's Republic of China}
\affiliation{Synergetic Innovation Center of Quantum Information and Quantum
  Physics, University of Science and Technology of China, Hefei, 230026,
  China}
\author{Lixin He}
\email{helx@ustc.edu.cn}
\affiliation{CAS Key Laboratory of Quantum Information, University of Science and
  Technology of China, Hefei, Anhui, 230026,  People's Republic of China}
\affiliation{Synergetic Innovation Center of Quantum Information and Quantum
  Physics, University of Science and Technology of China, Hefei, 230026, China}
\date{\today }

\pacs{71.10.-w, 75.10.Jm, 03.67.-a, 02.70.-c}

\begin{abstract}
The projected entangled pair states (PEPS)
methods have been proved to be powerful tools to solve the strongly correlated quantum many-body problems
in two-dimension.
However, due to the high computational scaling with the virtual bond dimension $D$, in a practical application PEPS
are often limited
to rather small bond dimensions, which may not be large
enough for some highly entangled systems, for instance, the frustrated systems.
The optimization of the ground state using imaginary time evolution method with simple update
scheme may go to a larger bond dimension. However, the accuracy of the rough approximation
to the environment of the local tensors is questionable.
Here, we demonstrate that combining the imaginary time evolution method with simple update, Monte Carlo sampling techniques and gradient optimization
will offer an efficient method to calculate the PEPS ground state.
By taking the advantages of massive parallel computing, we can study the quantum systems with larger bond dimensions
up to $D$=10 without resorting to any symmetry.
Benchmark tests of the method on the $J_1$-$J_2$ model give impressive accuracy compared with exact results.

\end{abstract}
\maketitle

\date{\today}

\section{Introduction}
Developing efficient algorithms to simulate strongly correlated quantum many-body systems is in the center of the modern condensed matter physics. In the context of strongly interacting systems, where the conventional perturbation theory fails, revealing their physical nature is mainly dependent on the numerical simulation methods, such as, exact diagonalization (ED), quantum Monte Carlo (QMC) method  and density matrix renormalization group (DMRG)~\cite{white92}. These numerical methods have been widely used in studying strongly correlated quantum systems and
have achieved great success. However, developing new efficient algorithms is still urgent, because of the limitations of the previous methods: e.g.,
ED encounters the so-called ``Exponential Wall''; QMC suffers from the notorious sign problem for fermionic and frustrated systems\cite{Troyer05}; and DMRG is limited to 1D or quasi-1D systems and does not work well for higher dimension systems\cite{Schollw11}.

Recently, inspired by the insight of quantum entanglement in the perspective of quantum information theory, the algorithms based on the tensor network states (TNS), particularly, matrix product states (MPS)\cite{vidal03,vidal04} and projected entangled pair states (PEPS)\cite{verstraete04,verstraete06}, which is a natural extension of MPS to higher dimensions that satisfies both area law\cite{verstraete08} and size consistency\cite{wangzhen13}, have been proved to be
powerful simulation methods to exploit the strongly correlated systems. The algorithms based on TNS offer great opportunities to solve some long standing two-dimensional problems.

Nevertheless, there are still some difficulties hindering the power of the simulation due to the complexity of PEPS and our limited computing capability. One of the major
difficulties is how to efficiently obtain the optimal PEPS with large bond dimension $D$. Generally, to make the obtained PEPS converge to the exact ground state of the system, the virtual bond dimension $D$ should be as large as possible. However, the computational cost
increases fast with the increasing $D$.
Originally, an imaginary time evolution method was implemented to
optimize the PEPS wave functions with the computational cost scales as O$(D^{10})$ for square lattice
with open boundary condition (OBC)~\cite{verstraete04,verstraete08}, because one needs to contract the whole tensor network to
calculate the environment of the local tensors.
Such high computational cost limits the bond dimension $D$ to quite small values (such as $D$=4 in Ref[\onlinecite{verstraete04}]), and as a result,
the application of the method is limited.
To reduce the calculation cost, an imaginary time evolution with simple update (SU) algorithm was proposed~\cite{simpleupdate}.
In this scheme, the environment of a tensor is approximated by products of some diagonal matrices, and therefore it
 substantially reduces the calculation cost in the update process to O$(D^5)$ by combining QR/LQ decompositions
 when dealing with nearest neighbor (NN) interactions~\cite{wang11}.
 However, the SU imaginary time evolution is a local optimization method,
 where the environment of the local tensor may be over simplified. Consequently, the optimized PEPS may not
 converge to the real ground state of the system with desired precision, especially for finite systems~\cite{lubasch14}.
Great effort has been  made to improve the results of SU~\cite{jordan08,orus09,orus15,wanglingcluster,singlelayer}. Recently a cluster update method\cite{lubasch14}, allowing a tradeoff between computation cost and precision, is proposed to improve the accuracy systematically by approximating the environment  of local tensors with different clusters of sizes. Based on the cluster update method, a full update (FU) method \cite{lubasch14,fullupdate14}, meaning taking the whole lattice into account, can significantly improve the accuracy from that of the SU. Unfortunately, the computational scaling of the FU
is still O$(D^{10})$, which prevents one from using larger $D$ in PEPS.

Besides the difficulty to efficiently find the ground state of a system, how to efficiently calculate the correlation functions is another one.
 A direct contraction method has a high computation scaling of O$(D^{10})$ for OBC systems
 (the same as the cost to get the environment of local tensors),
 so even if an accurate ground state with large $D$ is obtained, the time cost for computing
correlation functions may be still beyond our current capability.

To extend the ability to investigate TNS of larger $D$,
Monte Carlo (MC) sampling techniques have been introduced by Sandvik et.al ~\cite{sandvik07},
based on one dimensional MPS, and  Schuch et.al ~\cite{schuch08,cirac10}, based on string-bond states\cite{schuch08} in two-dimension systems,
which reduce dramatically the scaling of computational cost to the bond dimension $D$,
compared to the standard contraction method.
However, even though the MC sampling method has been used to calculate some physical quantities for a given PEPS~\cite{wang11},
it has not been applied to optimize PEPS wave functions themselves.

In this work, we demonstrate a global gradient optimization (GO) method
combined with the MC sampling technique to optimize the PEPS ground state.
During the process, the energy and energy gradients are calculated through the MC sampling technique which
dramatically reduces the scaling of the calculation to O$(MD^6)$, where $M$ is the sampling sweeps, significantly
less than that of the variational method and imaginary time evolution FU method with O$(D^{10})$.
After we obtain the optimized PEPS, we may calculate the correlation functions
through the MC sampling technique, with the same computational scaling.
This methods may further take the advantages of massive parallelization,
and therefore allow us to investigate strongly correlated quantum systems using
PEPS with much larger $D$.
We benchmark the method using the Heisenberg model and the $J_1$-$J_2$ model on a square lattice. We
calculate the ground state using the PEPS with bond dimension up to $D$=10.
Our results show that the method can give impressive accuracy that is significantly better than the SU, and even FU.

\section{Methods}
\label{sec:methods}

\begin{figure}
 \centering
 \includegraphics[width=3.2in]{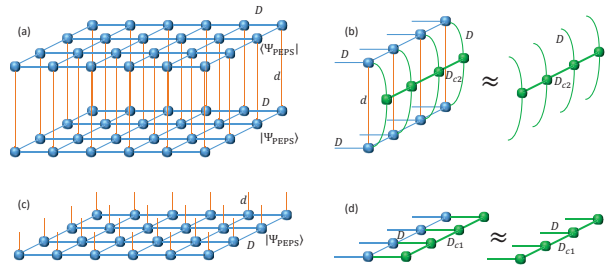}
 \caption{Contraction method for finite tensor network. (a) A double-layer tensor network composed of bra $\langle \Psi_{\rm PEPS}|$ and ket $|\Psi_{\rm PEPS}\rangle$ is needed to be contracted to obtain physical quantities by summing over the physical index  in the standard contraction scheme. (b) Boundary-MPO method is used to contract a double-layer tensor network with computational scaling O$(D^4D_{c2}^3+dD^6D_{c2}^2)$.(c) A single-layer tensor network $|\Psi_{\rm PEPS}\rangle$ with some give spin configuration $|S\rangle$ is needed to be contracted to obtain physical quantities in the MC scheme. (d)Boundary-MPS method is used to contract a single-layer tensor network with computational scaling O$(D^4D_{c1}^2)$.}
 \label{fig:contract}
\end{figure}

Considering a square lattice with $N\times M$ sites, and
$d$-dimensional local Hilbert space (which we call ``spin'' in this work)
whose bases are denoted as $|s_{i,j}\rangle$ on the site ($i$,$j$).
The PEPS wave function of this system can be written as\cite{verstraete04},
\begin{equation}
  |\Psi_{\rm PEPS}\rangle = \sum_{s_{1,1} \cdots s_{N,M}=1}^d { \rm Tr} (A_{1,1}^{s_{1,1}} A_{1,2}^{s_{1,2}}
  \cdots A_{N,M}^{s_{N,M}})|s_{1,1} \cdots s_{N,M} \rangle,
   \label{Eq:PEPS}
\end{equation}
shown in Fig.~\ref{fig:contract}(c),where $A^{s_{i,j}}_{i,j}$=$A_{i,j}(l,r,u,d,s_{i,j})$ is a five-index tensor located on site ($i$,$j$).
It has one physical index $s_{i,j}$ whose value is from $1$ to $d$
and four virtual indices $l,r,u,d$ corresponding to four nearest neighbors.
The dimension of each virtual bond is $D$, except for those on open edges,
whose dimensions are set to one.
The ``${\rm Tr}$'' denotes the contraction over all the virtual indices of the tensor network. The number of the parameters in the PEPS is determined by the bond dimension $D$. Some tricky many-body states (e.g., spin liquid) require large $D$ to give an accurate description.

PEPS provide systematically improvable variational wave functions to approximate
the exact many-particle states. Our goal is to optimize the PEPS wave functions and
obtain the physical quantities of the corresponding many-body ground states.
The contraction of the PEPS plays the central role and costs the dominant resource in our calculation. To calculate the environment of the local tensor during the optimization and the physical quantities for a given PEPS wave function, we need to contract the PEPS.
However, exactly contracting PEPS is NP-hard.~\cite{verstraete06_2}
The bond dimensions will grow exponentially with the number of the lines contracted
during the process. Therefore a truncation of the bond dimension
is necessary during the contraction~\cite{verstraete08}.
In the standard contraction methods to calculate energy or correlation functions,
one has to contract a double-layer tensor network with bond dimension $D^2$ composed of
 both bra $\langle \Psi_{\rm PEPS}|$ and ket $|\Psi_{\rm PEPS}\rangle$ by summing over the physical index, shown in Fig.~\ref{fig:contract}(a).  A boundary-MPO with bond dimension $D_{c2}$ approximation is introduced to avoid exponential growth of the bond dimension of the tensor network during contracting process\cite{lubasch14,fullupdate14}, shown in Fig.~\ref{fig:contract}(b). Therefore, even with this approximation, the computational scaling of the whole contraction is still O$(D^4D_{c2}^3+dD^6D_{c2}^2)$ for OBC . It has been shown, to get enough precision
the cut-off bond dimension $D_{c2}$ should be proportional to $D^2$
independent of the system size \cite{verstraete08,fullupdate14}, i.e., $D_{c2}\propto D^2$. Therefore,
the computational scaling to contract the whole PEPS is as high as O$(D^{10})$ for OBC.
The computational scaling for PBC is even higher with O$(D^{18})$ \cite{cirac10}.

To reduce the high scaling of the methods, Monte Carlo sampling techniques have been introduced
by several groups~\cite{sandvik07,schuch08,wang11}
in which the contraction over physical indices is replaced by the MC sampling over the ``spin configurations''.
In this algorithm, the energy is reexpressed as follows,
\begin{equation}
  E= \frac{\langle \Psi_{\rm PEPS} |H|\Psi_{\rm PEPS}\rangle}{\langle \Psi_{\rm PEPS} | \Psi_{\rm PEPS}\rangle}=\frac{1}{Z}\sum_{S}{W^{2}(S)E(S)} ~~,
 \end{equation}
with
\begin{equation}
E(S)=\sum_{S^{\prime}} \frac{W(S^{\prime})}{W(S)} \langle S^{\prime}|H|S\rangle \,.\nonumber
\end{equation}
Here $|S\rangle=|s_{1,1}s_{1,2}\cdots s_{N,M} \rangle$ is the spin configuration and
\begin{equation}
W(S)= { \rm Tr} (A_{1,1}^{s_{1,1}} A_{1,2}^{s_{1,2}}\cdots A_{N,M}^{s_{N,M}})\nonumber
\end{equation}
is the weight of the spin configuration.
$Z=\sum_{S}W^2(S)$ is the normalization factor.
The energy is evaluated through MC sampling according to the configuration weight $W^{2}(S)$.
Unlike the standard contraction method, in the MC scheme, the most time consuming part is to calculate
$W(S)$,
which is obtained by contracting a single-layer (instead of two layers in the original methods)
PEPS with bond dimension $D$ with fixed spin configurations, shown in Fig.~\ref{fig:contract}(c). Similar with boundary-MPO method, a boundary-MPS method is  used to approximately contract the single-layer tensor network, shown in Fig.~\ref{fig:contract}(d)\cite{verstraete08}. The scaling of this process is O$(D^4D_{c1}^2)$. Usually, $D_{c1} \sim 2D$ is enough for most problems,
therefore total computational scaling to calculate the energy is O$(D^6)$. We note that
the cut-off bond dimension $D_{c1}$$\sim$$D$ of the single-layer TNS in the MC sampling method
is corresponding to the cut-off bond dimension $D_{c2}$$\sim$$D^2$ in the double-layer PEPS in the original contraction methods.

The energy derivation with respect to the tensor element $A_{lrud}^{s_m}$ can also be evaluated by MC sampling as:
\begin{equation}
\frac{\partial E}{\partial A_{lrud}^{s_m}}=2\langle\Delta_{lrud}^{s_m}(S)E(S)\rangle-2\langle\Delta_{lrud}^{s_m}(S)\rangle \langle E(S)\rangle,
\end{equation}
where $s_m$ is the physical index of tensor $A$ located on site $m$,
and $\langle \cdots \rangle$ denotes the MC average. $\Delta_{lrud}^{s_m}$ is defined as
\begin{equation}
\Delta_{lrud}^{s_m}(S)=\frac{1}{W(S)}\frac{\partial W(S)}{\partial A_{lrud}^{s_m}}=\frac{1}{W(S)} B_{lrud}^{s_m}(S) ~~ ,
\end{equation}
where $B_{lrud}^{s_m}(S)$ is the element of
\begin{equation}
B^{s_m}(S)={\rm Tr} (A_1^{s_1} A_2^{s_2}\cdots A_{m-1}^{s_{m-1}}A_{m+1}^{s_{m+1}}\cdots A_N^{s_N} )\, ,
\end{equation}
which is nothing but a four-index tensor summing over all the indices of the single-layer network except those linked
with site $m$  on the fixed configuration $|S \rangle$. The scaling of calculating the energy gradient is also O$(D^6)$.

 Once we have the energy gradients, we can adopt the GO methods to calculate the PEPS ground state.
The PEPS energy function can be mapped to a classical mechanic system by treating tensor elements as generalized coordinates.
 The total energy can then be optimized via the steepest decent methods,
 or molecular dynamic methods,\cite{liu15}
 making full use of the energy gradients. By combining with the replica exchange method,
 local minima can be escaped efficiently,\cite{liu15}
 making it a powerful scheme to simulate the ground state of complex systems \cite{dong2016}.
 Another efficient way of optimizing the tensor is proposed by Sandvik et.al,~\cite{sandvik07}
 which uses only the sign of energy gradients. This
 is very useful when MC sampling sweeps is not very large, and the energy gradients are not very accurately calculated.

There are several advantages of the GO method.
Firstly, unlike the local optimization methods, the GO method update all tensors simultaneously
and also the noise in the gradient may help avoid the local minima, which has some similarity
to the simulated annealing technique.\cite{Harju1997}
Secondly, the MC sweeps can be easily and massively parallelized. Thirdly, it is easy to deal with the systems beyond nearest neighbor interactions, such as $J_1$-$J_2$ square Heisenberg model. In standard contraction methods, the long range interactions will dramatically increase the computation cost. For example, according to the contraction method mentioned above, when dealing with the next nearest neighbor terms in $J_1$-$J_2$ model with OBC, the calculation scaling can be as high as  O$(D^{12})$ if $D_{c2}\propto D^2$. The computation cost will increase rapidly with the range of the interaction.
However in our method, only the weight $W(S)$ of a single-layer tensor network needs to be contracted, and the long range interactions
 can be easily calculated in the ``spin'' representation, and therefore the computational cost is still O$(D^6)$,
 no matter what the range of the interaction is.

\section{Computational details}

To save the computational time, the imaginary time evolution method with SU,\cite{simpleupdate}
a fast local optimization method, is adopted to give a
good approximation to the exact ground state which serves as
the starting point for the further optimization~\cite{simpleupdate}.
The PEPS wave function is then optimized via the GO method.

When employing simple updating, the Hamiltonian will be split into several parts comprised of mutually commuting terms,
and Suzuki-Trotter expansion\cite{Suzuki90,Suzuki91}
is used to expand the evolution operator approximately. The tensors are updated site by site by singular value decomposition (SVD)
and the bond dimensions are truncated back to $D$. With the help of QR decomposition, the computational scaling of
 this process is O$(D^5)$ if only NN interactions are present,~\cite{wang11}
 and therefore, it is possible to use very large $D$ (e.g. $D \sim$ 20) in the SU method.
We perform imaginary time evolution starting with time step $d\tau$=0.01 until the tensors are converged, i.e., $\frac{|A_{lrud}^{s_m}(\tau+d\tau)-A_{lrud}^{s_m}(\tau)|}{|A_{lrud}^{s_m}(\tau)|}<10^{-6}$.
We then reduce the time step to $d\tau=0.001$, and keep optimizing the PEPS until
it is converged under the new time step.
When optimizing the PEPS with large bond dimension $D$, we do not directly start with random tensors.
Instead, we always  start with a random PEPS state with smallest $D$ (usually $D=2$), and the converged tensors are used
as the initial states for larger $D$. We gradually increase $D$ to the desired values.

With the PEPS optimized by SU, a GO is adopted with tensor elements varying as follows:
\begin{equation}
  A_{lrud}^{s_m}(n+1)=  A_{lrud}^{s_m}(n)-p\cdot\delta t(n)\cdot {\rm{sign}}(\frac{\partial E}{\partial A_{lrud}^{s_m}}) ~~,
\end{equation}
where $p$ is a random real number ranged from $0$ to $1$ for each $A_{lrud}^{s_m}$, which can help  avoid trapping at local minima,
and $\delta t(n)$ is the step length.
We would like to point out that only  the correct signs of energy derivations ($\frac{\partial E}{\partial A_{lrud}^{s_m}}$) are needed instead of absolute values in this method.~\cite{sandvik07}
Alternatively, one can also use the molecular dynamics method described in Ref.[\onlinecite{liu15}] to optimize the PEPS.

\begin{figure}[bt]
 \centering
 \includegraphics[width=2.5in]{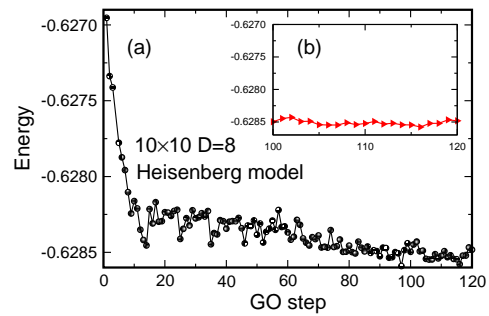}
 \caption{(Color online) GO for Heisenberg model on a $10\times10$ square lattice using PEPS with $D$=8. (a) In the first 50 GO steps, $\delta t$ is set
 to 0.005. In the next 50 steps,  $\delta t$ decreases slowly from 0.005 to 0.001, and in the last 20 steps, $\delta t$=0.001.
 (b) The energy variation in the last 20 steps.  }
 \label{fig:GOevolution}
 \end{figure}

Figure~\ref{fig:GOevolution} depicts a typical GO process for the Heisenberg model on a $10 \times 10$ square lattice with $D$=8.
We start from an approximate ground state obtained by SU. In the first 50 GO steps, we set $\delta t(n)=0.005$, and gradually
increase MC sampling number from $M$=50000 to $M$=100000.
We reduce $\delta t$ slowly from 0.005 to 0.001 in the next 50 GO steps using $\delta t(n+1)=\delta t(n)*0.968$, and gradually increase $M$. In the last
20 steps, we use fixed $\delta t$=0.001, and the maximum $M$=500000.
It takes about 25 minutes for each GO step using 500 Intel E5-2860 cores for the maximum $M$. Our tests show that the combined SU method
and GO method is a very robust method to optimize the PEPS.
We have also tried the GO optimization starting form a random PEPS, which turns out to be very expensive and often trapped at local minima.
Of course, we can start the GO optimization using previously optimized PEPS of smaller $D$, which is also a good starting point.

In the following section, we will compare the results obtained from our method with the results obtained with the previous SU and FU methods for the $J_1$-$J_2$ model on square lattices.
The full methods described in this paper are implemented using an in house Fortran2003 library~\cite{TensorLib} designed for tensor network states methods.

\begin{table*}[t]\footnotesize
\caption {Comparison of ground state energies of the $J_1$-$J_2$ model on the square lattices
obtained by the simple update (SU) imaginary time evolution method and gradient optimization (GO) method to the exact results.
For the 4$\times$4 and 4$\times$6 lattices, the exact results are obtained by exact diagonlization method,
whereas for the 10$\times$10 lattice,
the exact result refers to the one obtained from QMC simulations.~\cite{MCHeisenberg97}}
\centering
	\begin{tabular}{ccccccccccc}
\hline\hline
   \multicolumn{1}{c}{\multirow{2}{*}{$D$}}
   &\multicolumn{2}{c} {\multirow{1}{*}{$4\times4 ~~ J_2/J_1=0.0$}}
    &\multicolumn{2}{c} {\multirow{1}{*}{$4\times6 ~~ J_2/J_1=0.0$}}
   &\multicolumn{2}{c} {\multirow{1}{*}{$4\times6 ~~ J_2/J_1=0.50$}}
   &\multicolumn{2}{c}{\multirow{1}{*}{$4\times6 ~~ J_2/J_1=0.56$}}
     &\multicolumn{2}{c}{\multirow{1}{*}{$10\times10 ~~ J_2/J_1=0.0$}}\\
   & SU & GO & SU & GO & SU & GO & SU & GO  & SU & GO  \\
   \hline
 		2 & -0.54557(3)&-0.570872(2) &-0.57146(5) &-0.581942(7) &-0.45616(1) & -0.466301(8) & -0.43308(1) & -0.451364(2) & -0.61281(1) & -0.617795(4)\\
 		3 & -0.55481(1)&-0.573625(2) &-0.57703(1) &-0.586866(4) &-0.46824(2) & -0.473510(2) & -0.45252(3) & -0.461808(6) & -0.61846(3) & -0.624187(2)\\
		4 & -0.56317(2)&-0.574284(1) &-0.58122(2) &-0.588643(1) &-0.46980(3) & -0.474200(9) & -0.45287(4) & -0.462807(1) & -0.62412(2) & -0.627894(1)\\
		5 & -0.56660(1)&-0.574312(1) &-0.58300(1) &-0.588701(2) &-0.47168(1) & -0.474328(7) &-0.45922(8)  & -0.463377(1) & -0.62514(1) & -0.628412(9)\\
 		6 & -0.56714(4)&-0.574316(2) &-0.58311(2) &-0.588702(2) &-0.47265(4) & -0.474346(9) & -0.46024(7) & -0.463438(1) & -0.62541(2) & -0.628448(1)\\
 		7 & -0.56715(2)&-0.574318(1) &-0.58337(3) &-0.588705(2) &-0.47283(1) & -0.474356(2) & -0.46057(1) & -0.463441(2) & -0.62543(5) & -0.628488(2)\\
       	8 & -0.56725(6)&-0.574319(2) &-0.58341(6) &-0.588707(1) &-0.47318(3) & -0.474358(1) & -0.46130(3) & -0.463446(1) & -0.62566(6) & -0.628507(1)\\
             9 & -0.56727(3)&-0.574321(1)&-0.58358(1) & -0.588712(5) &-0.47364(2) & -0.474362(1) & -0.46213(4) & -0.463475(2) & -0.62570(2) & -0.628566(1)\\
            10& -0.56942(5)&-0.574323(1) &-0.58470(1) &-0.588713(3) &-0.47379(1) & -0.474365(1) & -0.46214(3) & -0.463476(1) & -0.62611(1) & -0.628601(2)\\   \hline
 	 Exact&\multicolumn{2}{c}{ -0.57432544}
     & \multicolumn{2}{c}{-0.58871445}
      &\multicolumn{2}{c}{ -0.47437906}
      & \multicolumn{2}{c}{-0.46350353}
      & \multicolumn{2}{c}{-0.628656(2) (MC)~\cite{MCHeisenberg97}} \\
    \hline\hline
	\end{tabular}
\label{tab:Eenrgybenchmark}	
\end{table*}

\begin{figure}[htb]
 \centering
 \includegraphics[width=2.5in]{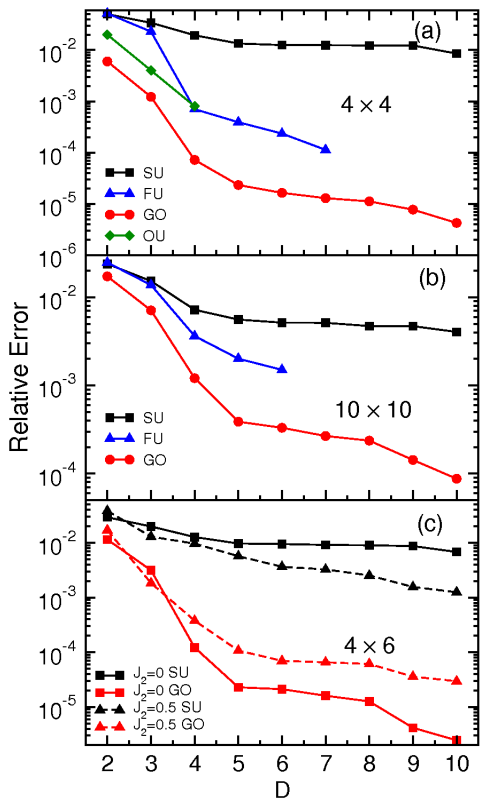}
 \caption{(Color online) Benchmark testes on the total energies of the spin-${1 \over 2}$ Heisenberg antiferromagnetic model on square lattices
  as functions of PEPS bond dimension $D$. The energies are obtained by different optimization methods including
  the original update (OU), simple update (SU), full update (FU) and gradient optimization (GO) methods,
  on different lattice sizes and $J_2$ parameters: (a) a 4$\times$4 lattice with $J_2$=0; (b) a 10$\times$10 lattice with $J_2$=0;  and (c) a 4$\times$6 lattice with $J_2$=0.5. The error bars are too small to show.}
 \label{fig:Heisenberg}
 \end{figure}

\section{Benchmark results}


We benchmark the method using a typical two dimensional frustrated spin-$1/2$ Heisenberg model,
namely the $J_1$-$J_2$ model on a square lattice. The Hamiltonian of the model reads,
\begin{equation}
  H = J_1 \sum_{\langle i,j\rangle} {\bf S}_i \cdot {\bf S}_j  + J_2
  \sum_{\langle \langle i,j \rangle \rangle} {\bf S}_i \cdot {\bf S}_j\ .
\end{equation}
The spin operators obey ${\bf S}_i \cdot {\bf S}_i=S(S+1)$=3/4, whereas
$\langle i,j \rangle$ and $\langle \langle i,j \rangle \rangle$
denote the nearest and next-nearest neighbor spin pairs, respectively,
on the square lattice. Without loss of generality, we set $J_1$=1 in all calculations.
When $J_2>$0, there are frustrated interactions, between the nearest and next-nearest
spin pairs.
$J_1$-$J_2$ model has been extensively studied, because
it has rich physics and is an interesting model whose ground state may be a
spin liquid or a plaquette valence-bond  state near $J_2/J_1=0.5$\cite{Jiang12,Hu13,wang16,Gong14}.

\begin{figure}[b]
 \centering
 \includegraphics[width=2.5in]{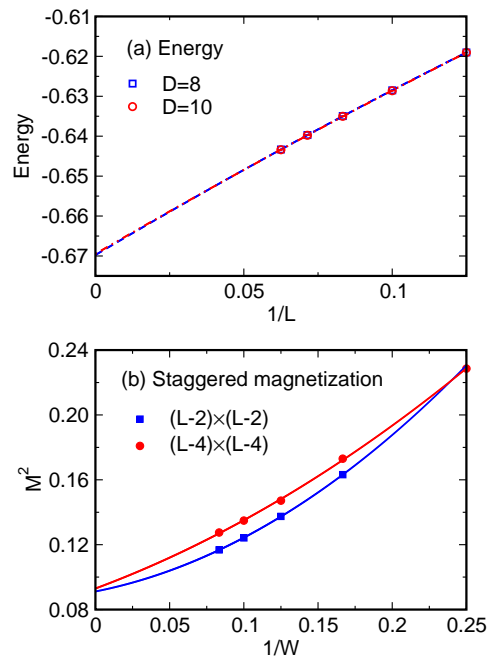}
 \caption{(Color online) (a) The ground energies of PEPS $D$=8 and $D=10$, and (b) the staggered magnetization $M^2$ of
the Heisenberg model, calculated by PEPS with $D$=8, on the $L=8,10,12,14,16$ square lattices.
The staggered magnetization is calculated on the central $W\times W$ region, with $W$=$(L-2)$ and $(L-4)$ respectively, to reduce the boundary effect.
}

 \label{fig:Ms-energy}
 \end{figure}

\begin{table*}[tb]
\caption{Comparison of the ground state energies of the Heisenberg model calculated by PEPS
with available exact results for $L=6,~8,~10,~12,~14,~16$. The PEPS are optimized via the GO method,
with $D$=8, and $D_c$=16. The exact result for
$L$=6 is taken from Ref.[\onlinecite{xiang2016}] obtained by DMRG with keeping 4096 states.
All other values are taken from  Ref.[\onlinecite{fullupdate14}] which are obtained by QMC method~\cite{ALPS1,ALPS2,ALPS3}.}
\centering
\begin{tabular}{c|ccccccc}
\hline\hline
\multicolumn{1}{c|}{$L$} &\multicolumn{1}{c}{6} &\multicolumn{1}{c}{8 }&\multicolumn{1}{c}{10}
&\multicolumn{1}{c}{12} &\multicolumn{1}{c}{14} &\multicolumn{1}{c}{16} &\multicolumn{1}{c}{$\infty$} \\  \hline
$D$=8   & -0.603523(1) & -0.619013(2) & -0.628507(1) & -0.634958(1)  & -0.639697(5) & -0.643330(5)  & -0.66977(25) \\ \hline
$D$=10  & -0.603535(1) & -0.619033(3)& -0.628601(2) & -0.635025(3) & -0.639764(3) & -0.643391(3)  & -0.66948(42) \\ \hline
Exact  & -0.6035218\cite{xiang2016}  & -- & -0.628656(2)\cite{fullupdate14} & --  & -0.639939(2)\cite{fullupdate14}& -0.643531(2)\cite{fullupdate14}&-0.6694437(5)\cite{MCHeisenberg97} \\ \hline\hline
\end{tabular}
\label{tab:HeisenbergEenrgy}	
\end{table*}

We first test our method on the simple Heisenberg model, with $J_2$=0.
In Fig.~\ref{fig:Heisenberg}(a),
we compare the results obtained from GO on the 4$\times$4 lattice
to those obtained by imaginary time evolution with the original update (OU)
algorithm\cite{verstraete04} proposed by Verstraete et. al.,
SU\cite{simpleupdate} and FU methods\cite{lubasch14,fullupdate14}
as functions of bond dimension $D$. The OU and FU results are taken from Ref.[\onlinecite{verstraete04}] and Ref.[\onlinecite{fullupdate14}] respectively.
We calculate the relative errors of these methods, defined as $|E-E_{\rm ex}|/|E_{\rm ex}|$, where the exact results $E_{\rm ex}$
are obtained from the exact diagonlization method. The OU method
is very accurate and the relative error can be reduced to 10$^{-3}$ even for $D$=4.
However, the scaling of $D$ in the OU method is too high, and
therefore, it is very difficult to use larger $D$ in the computations.
 The SU (black squares) permits a larger bond dimension $D$, but it gives rather large error,
 approximately 2\% at $D$=4, when compared to $E_{\rm ex}$, which does not improve much by further increasing $D$. Therefore the accuracy of SU may not
be enough for some problems, especially when there are competing phases, and simply increasing $D$ does not solve the problem.
The recent developed FU method (blue triangles)~\cite{fullupdate14} can achieve similar accuracy
 to OU at $D$=4, but is less computationally cost~\cite{lubasch14},
 and therefore is affordable for larger $D$.
 As $D$ increases from 4 to $D$=7, the relative error reduces from  10$^{-3}$  to 10$^{-4}$.
The relative errors of the total energy using the GO method are shown as red dots in Fig.~\ref{fig:Heisenberg}(a).
These results are systematically better
than those of the SU and FU methods.
With lower scaling to $D$, we can further use $D$=10,
and the relative error is reduced to approximately 4.2$\times$10$^{-6}$.
Detailed results of the total energies calculated by SU and GO are compared in
Table~\ref{tab:Eenrgybenchmark} with the exact values.
In all these calculations, we use $D_c$=2$D$, which converges  the results very well, as
discussed in Appendix.~\ref{sec:bond}.

For the 10$\times$10 lattice, the exact diagonlization method is
not applicable. We therefore compare the relative error of the total energies
obtained by SU, FU and GO to the available QMC results \cite{fullupdate14} in Fig.~\ref{fig:Heisenberg}(b).
The relative errors of SU show very similar behaviors as in the small size lattice.
The error is approximately 10$^{-2}$ at $D$=8, which changes only a little for up to $D$=10.
The best available results of FU in the literature is $D$=6. As we see, the
FU method greatly improves the results of SU method, and at $D$=6, the error of FU is approximately 2$\times$10$^{-3}$.
Again the results of GO is even better than FU for all bond dimension $D$,
and at $D$=10, the error of GO relative to QMC reduces to 8.7$\times$10$^{-5}$.

The total energies per site calculated from GO with $D$=8, $D$=10 are shown in
Fig.~\ref{fig:Ms-energy}(a),
and are compared with the available
exact diagonlization or QMC results for $L$=6, 8, 10, 12, 14, 16.
in Table II. The agreement between the PEPS results
and the exact results are remarkable.
By extrapolating the energies using a second-order polynomial fitting,
we obtain the ground state energies of the Heisenberg model $E$=-0.66977(25) and  $E$=-0.66948(42)
per site in the thermodynamic limit for $D$=8 and $D$=10 respectively by fitting the results of $L$=8, 10, 12, 14, 16,
which are in excellent agreement to
the QMC result $E$=-0.669437(5)~\cite{MCHeisenberg97}.

We also calculate the spin-spin correlation function $m^2_s({\bf k})=\frac{1}{N^2}\sum_{ij}{\langle{\bf S}_i \cdot {\bf S}_j}\rangle {e}^{i {\bf k}\cdot({\bf r}_i-{\bf r}_j)}$, where $N$ is the total number of spins included in the summation, using the ground state wave functions optimized from GO method.  The ground state of
Heisenberg model has an AFM order, with staggered magnization
$M^2$= $m^2_s(k_x=\pi, k_y=\pi)$.
To reduce the boundary effects, we restrict our summation to the central lattice with bulk size\cite{2dDMRG12}  $(L-2)$$\times$$(L-2)$ to obtain $M_1^2(L)$ and  $(L-4)$$\times$$(L-4)$ to obtain $M_2^2(L)$ , where $N=(L-2)^2$ and $N=(L-4)^2$ correspondingly.
We calculate $M^2$ on the lattice with different size $L$=8,~10,~12,~14,~16. The results are shown in Fig.~\ref{fig:Ms-energy}(b).
We extrapolate $M^2(L)$  to the thermodynamic limit $L\rightarrow \infty$ using a second-order polynomial fitting and
obtain $M_1^2(\infty)=0.091\pm0.001$ and $M_2^2(\infty)=0.093\pm0.002$, which gives $M_1(\infty)$=0.302 and $M_2(\infty)$=0.305,
both are in excellent agreement
to the best known numerical value of $M(\infty)=0.307$ by QMC simulations\cite{MCHeisenberg97}.
It is remarkable that the staggered magnization $M_1$ and $M_2$
are very close to each other as $L\rightarrow \infty$, which implies that the boundary effects can be effectively reduced by coping with the central
bulk regions for OBC systems\cite{2dDMRG12}. By comparing the energy and staggered magnetization in the thermodynamic limit, we find PEPS with $D$=8 is enough
 to capture the correct physics for Heisenberg model, if the PEPS wave functions are fully optimized.
The calculation of the staggered magnization is very expensive, since one needs
to calculate $\langle s_i \cdot s_j \rangle$ for all possible $(i,j)$ pairs with extremely large MC sampling numbers. For the 14$\times$14
lattice, and $D$=8, it takes 45000 CPU core-hours to get the $M_1^2(L)$ for $M$=5000000.

When $J_2 \neq$0, there are frustrated NNN spin interactions, where the standard QMC methods suffer from the notorious negative sign problem\cite{Troyer05}. In these cases,
the TNS methods show great advantages. It is widely believed, near $J_2 \sim$ 0.5, where the frustration is strongest, the ground state
of the $J_1$-$J_2$ model might be a highly entangled spin liquid state \cite{Jiang12,Hu13,wang16} or plaquette valence-bond state \cite{Gong14},
which presents a great challenge to all available numerical methods.
Fig.~\ref{fig:Heisenberg}(c) shows the energies on a 4$\times$6 lattice, obtained by SU and
GO with different $D$ at $J_2$=0 and 0.5.
Compared with the exact diagnolization results,
the relative errors of the energies obtained from GO at $D=10$
are 2.5$\times$10$^{-6}$ for $J_2$=0 and 3.0$\times$10$^{-5}$ for $J_2$=0.5 respectively, suggesting that  the GO method is
also good for the non-trivial problems and indicating PEPS are good variational wave functions for the frustrated $J_1$-$J_2$ model.

\begin{table}[htb]
\caption{Comparison of some currently used optimization methods for finite PEPS on square Heisenberg model
in terms of computational scaling, maximal bond dimension $D$, maximal lattice size $L$ and  relative error achieved
with respect to the best available results. The SU and MC+GO results are based on current calculations.}
\centering
\begin{tabular}{c|cccc}
\hline\hline
\multicolumn{1}{c|}{method} &\multicolumn{1}{c}{OU\cite{verstraete04,Murg2009}} &\multicolumn{1}{c}{SU }&\multicolumn{1}{c}{FU\cite{fullupdate14}}&\multicolumn{1}{c}{MC+GO}\\  \hline
scaling  & O$(D^{10})$&O$(D^{5})$ & O$(D^{10})$ & O($MD^{6})$  \\ \hline
max $D$  & 4 &  $\ge10$ & 6 & $10 $ \\ \hline
max $L$  & 14& $\ge16$ & 14 & $16$ \\ \hline
relative error  & $\sim10^{-3}$ & $\sim10^{-2}$  & $\sim10^{-3}$  &  $\sim10^{-4}$ \\ \hline \hline
\end{tabular}
\label{tab:difference}	
\end{table}

In Table~\ref{tab:difference}, we compare some of the currently used optimization methods for PEPS.
The results suggest that the combined MC and GO method can afford quite large $D$ and size $L$ with impressive accuracy.
Therefore the method is very promising for future applications using PEPS.
When we finalize the work, we get to know that variational optimization methods including the gradient method
\cite{corboz16,verstraete16}, have been applied to
optimize infinite PEPS wave functions where the gradients are
calculated via direct contractions.
It has also been shown that variational results are better than the best known full update results, consistent with our findings.

\section{summary}

We have demonstrated that a gradient optimization method,
combined with Monte Carlo sampling method and imaginary time evolution simple
update method, offers an efficient algorithm to optimize PEPS ground state
and calculate the correlation functions.
Benchmark tests on $J_1$-$J_2$ model show that
the method can give impressive accuracy that is significantly better than the simple update method.
By taking the advantages of massive parallelization, the method potentially can afford much larger bond dimension,
which is crucial to investigate highly entangled physical systems with topological orders and fermionic systems.
A future direction is to impose symmetry to the tensors, which is a promising route to
 boost the bond dimension one can afford.

\acknowledgements

This work was funded by the Chinese National Science Foundation (Grant number 11374275, 11474267),
the National Key Research and Development Program of China (Grants No. 2016YFB0201202).
The numerical calculations have been done on the USTC HPC facilities.

\begin{figure}[t]
 \centering
 \includegraphics[width=3.0in]{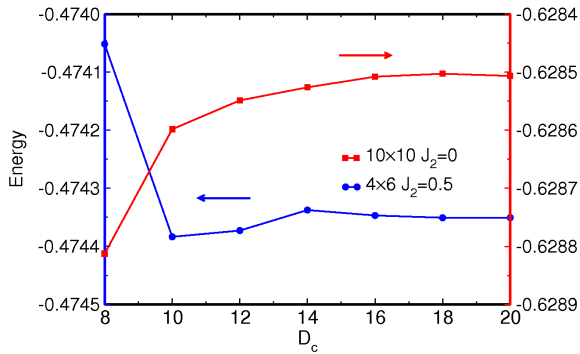}
 \caption{(Color online) The convergence of the ground state energies as functions of bond dimension cut-off $D_c$
 at $D$=8, for a Heisenberg model on a 10$\times$10 lattice and a $J_1$-$J_2$ model with $J_2$=0.5 on a 4$\times$6 lattice.
 The MC sampling error is order of $10^{-6}$.}
 \label{fig: converge}
 \end{figure}

\vskip2cm
\appendix

\section{Bond dimension truncation}
\label{sec:bond}

When contracting a two-dimensional PEPS, a truncation has been made
to avoid the exponential growth of the bond dimensions during the process.
The cut-off bond dimension $D_c$ may affect the final results.
We test the effects of $D_c$ on the convergence of the total energy. The typical results
for  Heisenberg model on a 10$\times$10 lattice and  $J_1$-$J_2$ model, with $J_2$=0.5,
on a 4$\times$6 lattice are shown in Fig.~\ref{fig: converge}.
For a fixed bond dimension $D$=8, we test $D_c$=8 to 20.
For the 10$\times$10 Heisenberg model, $D_c=16$ can converge the energy with an absolute error 8$\times$10$^{-6}$, comparing with $D_c=20$.
For the $J_1$-$J_2$ model on a 4$\times$6 lattice, the convergence of the total energy with $D_c$ shows
similar behavior.
Considering the balance between accuracy and computational cost,
$D_c=2D$ is adopted throughout
this paper to calculate expectable values of the observable including energy and correlation functions.


\end{document}